\documentclass[twocolumn,prl,aps,showpacs]{revtex4}
\usepackage{xspace,amsmath,amsfonts,amsthm,amssymb,amsbsy}
\usepackage{graphicx}
\usepackage{epstopdf}

\newcommand{\sq}[1]{\left[ {#1} \right]}
\newcommand{\tr}[1]{{\textrm {Tr}}\sq{#1}}
\newcommand{\smallfrac}[2]{\mbox{$\frac{#1}{#2}$}}
\newcommand{\half}{\smallfrac{1}{2}}
\newcommand{\bra}[1]{\langle{#1}|}
\newcommand{\ket}[1]{|{#1}\rangle}

\newcommand{\op}[2]{\ket{#1}\bra{#2}}
\newcommand{\enavg}[1]{\mathrm{E}\sq{#1}}
\newcommand{\expt}[1]{\langle{#1}\rangle}
\newcommand{\dg}{^\dagger}
\newcommand{\D}[1]{{\cal D}\sq{#1}}

\newcommand{\Hc}[1]{{\cal H}\sq{#1}}

\newcommand{\beq}{\begin{equation}} 
\newcommand{\eeq}{\end{equation}}
\newcommand{\bqa}{\begin{eqnarray}} 
\newcommand{\eqa}{\end{eqnarray}}
\newcommand{\nn}{\nonumber} 

\newcommand{\erf}[1]{Eq.~(\ref{#1})}
\newcommand{\frf}[1]{Fig.~\ref{#1}}

\newcommand{\hei}{Heisenberg }

\newcommand{\ito}{It\^o }

\newcommand{\cu}[1]{\left\{ {#1} \right\}}

\newcommand{\an}[1]{\left\langle{#1}\right\rangle}

\renewcommand{\section}[1]{{\em #1}.---}

\begin{document}

\newtheorem{theo}{Theorem}
\newtheorem{lemma}{Lemma}

\title{Replacing Quantum Feedback with Open-Loop Control and Quantum Filtering}

\author{Joshua Combes}
\author{Howard M. Wiseman}
\author{A. J. Scott}
\affiliation{Centre for Quantum Computer Technology, Centre for Quantum Dynamics, Griffith University, Nathan 4111, Australia}

\begin{abstract}
Feedback control protocols can stabilize and enhance the operation of quantum devices, however, unavoidable delays in the feedback loop adversely affect their performance. We introduce a quantum control methodology, combining open-loop control with quantum filtering, which is not constrained by feedback delays. For the problems studied (rapid purification and rapid measurement) we  analytically derive lower bounds on the control performance that are comparable with the best corresponding bounds for feedback protocols.  
\end{abstract}


\maketitle
Open-loop control and feedback control are generally perceived to be distinct but complementary control methodologies \cite{WisMil09}. Open-loop control is usually independent of measurement while feedback control, of course, involves measurement. Both types of control are powerful experimental and theoretical tools in classical and quantum contexts. In the quantum setting, open-loop methods have, for example, allowed: the calculation of bounds on gate complexity in quantum logic \cite{NieDowGu06sci} and optimal gate synthesis \cite{Gatesyn}; and decoherence suppression by bang-bang control and dynamical decoupling \cite{openloop2}. Equally impressive progress has been made in feedback control theory: the preparation of squeezed states of light and spin \cite{Sque}; stabilization of pure and entangled states \cite{Stable}; continuous error correction \cite{errorz}; and precision metrology and hypothesis testing \cite{adaptme}. Excitingly, these and other theoretical advances in quantum control are being realized in experiments \cite{exp,exp2}.

In spite of these successes there are significant challenges facing the field, in particular, those in quantum feedback control stemming from delays in the feedback loop. When such delays are larger than the relevant system timescales, the controllability of the system is severely degraded \cite{fbdelay}. In \frf{Fig1} we depict important sources of delays for each component in the feedback loop. Typical dynamical timescales ($\tau_{\mathrm{sys}}$) in the relevant quantum systems are $10^{-9}$ to $10^{-3}$s \cite{MabKha05}. In some experiments the reciprocal detector bandwidth ($\tau_{\mathrm{det}}$) and electronic delays dominate the total effective feedback delay $\tau_{\rm del}$ \cite{exp2}; in others, the time taken to calculate the conditional state ($\tau_{\mathrm{fil}}$) and the optimal control ($\tau_{\mathrm{ctrl}}$) will be most important. Currently, many atomic feedback experiments are limited by the response time of the actuator ($\tau_{\mathrm{act}}$), such as usually an electro-optic-modulator~\cite{StoArmMab02}, but  in solid state systems, $\tau_{\mathrm{act}}$ is typically small compared to the other time scales~\cite{driely}. In the near future, as actuator bandwidths improve, the field of quantum feedback control will be placed in an awkward position: even if $\tau_{\mathrm{act}}\ll \tau_{\mathrm{sys}}$, it may be impossible to perform effective feedback control because $\tau_{\rm del}\gg \tau_{\mathrm{sys}}$.

\begin{figure}
\leavevmode \includegraphics[width=0.8\hsize]{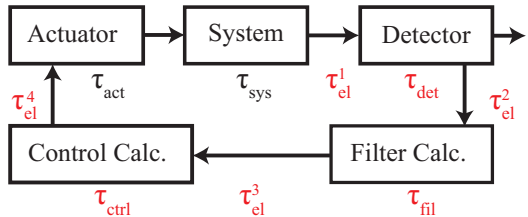}
\caption{(color online). A schematic of important time scales in a feedback loop. For devices, the times depicted are the reciprocals of the component bandwidths. The $\tau_{\mathrm{el}}^i$'s are electronic delays between devices. The total effective delay is the sum of all the red terms: $\tau_{\mathrm{del}}\equiv \sum_i \tau_{\rm el}^i + \tau_{\mathrm{det}}+\tau_{\mathrm{fil}}+\tau_{\mathrm{ctrl}}$.   
  \label{Fig1} }
\end{figure}

In this Rapid Communication we introduce a type of quantum control which is a hybrid of techniques from open-loop and feedback control, and that is not limited by feedback delays (or at least is considerably less so). The control objectives we consider include rapid state purification \cite{ComJac0601}, rapid state measurement \cite{ComWisJac08,WisRal06},  and rapid state preparation \cite{Jac04WisBou08}. Our proposal is to apply open-loop control to the system over some time interval $[0,T]$ (with $T\gg \tau_{\rm sys}$), while also continuously monitoring it. With the resulting measurement record, one may perform filtering (calculating the conditioned state \cite{filterref, WisMil09}) either in parallel to the strategy, or `offline' (after time $T$). Offline filtering  is suitable for objectives like rapid measurement and purification, which require no feedback at all. Filtering in parallel (requiring $\tau_{\mathrm{del}} \ll T$) may be necessary for the objective of rapid cooling or state preparation, which in our scheme entails performing a single control step after time $T$. In all cases, by combining continuous conditioning and open-loop control we obtain results comparable to those previously derived for measurement-based quantum feedback. Our methodology is distinct from control based on: quantum back-action~\cite{qfb_using_qba}, the quantum anti-Zeno effect~\cite{antizenocontrol}, learning~\cite{lern}, or coherent feedback~\cite{coherentcontrol}. 

This Rapid Communication is organized as follows. First, we briefly review continuous measurements and quantum filtering. Next we examine the problem of rapid purification under a random-unitary control strategy. We then consider the rapid measurement control problem, with the unitary controls restricted to the permutation group. In both cases we prove analytically that the scaling of the speed with the dimension of the system is the same as in the feedback-control strategies of Refs.~\cite{ComJac0601,ComWisJac08}. We numerically examine the effect of varying $\delta t$, the time between control pulses, and find that a speed-up can persist even with $\delta t \sim \tau_{\rm sys}$. We conclude with a discussion of some implications for quantum control, and open questions.

\section{Quantum Filtering} 
Consider a $D$-dimensional quantum system undergoing a continuous measurement of an observable $X$.  The change to our state of knowledge of an individual system, $\rho$, conditioned on the result of the measurement in an infinitesimal interval is described by the stochastic master equation (SME) ~\cite{CMreview,WisMil09} 
\begin{eqnarray}\label{SME}
 d\rho[t;X]= 2\gamma \, dt\, \D{X}\rho(t) +\sqrt{2\gamma}\,dw(t)\,\Hc{X}\rho(t),  
  \end{eqnarray}
where $\D{A} \rho \equiv A\rho A\dg -\half (A\dg A \rho + \rho A\dg A)$ and $\Hc{A} \rho \equiv  A\rho +\rho A\dg - \tr{(A\dg+ A )\rho}\rho$. We are working in a frame that removes any Hamiltonian evolution. The {\em measurement strength}, $\gamma$, determines the rate at which information is extracted. The measurement  result in the interval $[t,t+dt)$ is  $dR = \sqrt{4\gamma}\langle X(t) \rangle dt + dw(t)$, where $dw$ is a Wiener process and $\expt{X(t)} = \tr{X\rho(t)}$. Without loss of generality we take $X$ to be traceless. We can do this because \erf{SME} is invariant under $X \to X+\lambda I$ for $\lambda \in \mathbb{R}$.

We wish to combine this conditional evolution with an open-loop  control strategy which comprises applying a sequence of unitaries $U_q$ at times  $t_q=t_0+q\delta t$. Such a ``stroboscopic'' strategy, requiring arbitrarily strong control Hamiltonians, is also used in bang-bang control \cite{openloop2}, for example. This modifies the above quantum  filtering equation as follows. Define $U(t)$ for any $t \geq t_0$ as $U_q\ldots U_2U_1$ for $t_{q}\leq t<t_{q+1}$. Working in the \hei picture with respect to the control unitary, the monitored observable at time $t$ is rotated from $\check X(t_0)=X$ to $\check{X}(t)=U\dg(t){X}U(t)$. Thus the conditional increment of the system in interval $[t,t+dt)$ is simply $d\rho[t;\check{X}(t)]$. To use the information in the measurement record, the SME must be integrated as part of the experiment, but as noted above, this can happen `off-line'.

\section{First application: Rapid Purification} Pure states are necessary for many quantum information protocols. Motivated by this, Jacobs \cite{Jac0303}
introduced the following control goal: reducing the average of $L(t)=1-\tr{\rho(t)^2}$ (the impurity of the conditional state), to a fixed small value in the minimum time, using arbitrary Hamiltonian control of the system. In Ref. \cite{ComJac0601} this was studied for a $D$-dimensional system with $X$, the monitored observable, being $J_z$. It was shown that, by applying time-dependent controls to ensure that the conditional state is always unbiased with respect to the measurement basis, the rate of purification can be increased by a factor of at least $\smallfrac{2}{3}(D+1)$ over the no-control case. This control of course requires a feedback loop, as it depends on the conditional state, and hence may be difficult or impossible to implement because of feedback delays. Here we find an open-loop strategy that works almost as well: applying random unitaries. The intuition as to why this works is as follows. The feedback of Ref.~\cite{ComJac0601} maximizes the coherences of $\rho(t)$ in the measurement basis, and applying random unitaries also creates such coherences. By contrast, in the no-control case,  the coherences remain zero (or decay exponentially if they are initially non-zero).


To calculate analytically how the continuous measurement with random-unitary controls work, we make a simplifying approximation, that $\delta t \ll \gamma^{-1} = \tau_{\rm sys}$. That is, the controls are applied frequently on the characteristic system evolution time. This enables us to treat the evolution from $t_k$ to $t_{k+1}$ as an infinitesimal change described by \erf{SME} with $X \to \check X(t_k)$. Dropping the discrete time indices, we simply calculate the infinitesimal increment $dL$ from \erf{SME} with $X\to\check X(t) = U\dg(t)XU(t)$, assuming $U(t)$ to be a different random unitary at every time. Using the \ito calculus, we find $\enavg{dL}$, the change in the impurity, averaged over the measurement noise $dw$, to be 
\begin{eqnarray}\label{dlmpurity}
\nonumber \enavg{dL(\rho, \check{X})}&=& - 8\gamma \{ \tr{\check{X}\rho \check{X}\rho}-2\tr{\check{X}\rho}\tr{\check{X}\rho^2}\\
&&\phantom{ - 8\gamma dt \{}+\tr{\check{X}\rho}^2\tr{\rho^2}\}dt. 
\end{eqnarray}
Since $U(t)$ is drawn at random from the unitary group ${\mathfrak U}(D)$, we perform a further average over that group: 
\begin{eqnarray}\label{Uavg_dlmpurity}
\mathbb{E}[\enavg{dL(\rho)}]\equiv\int _{{\mathfrak U}(D)} dU\, \enavg{dL(\rho, U\dg X U)},
\end{eqnarray}
where $dU$ is the Haar probability measure on ${\mathfrak U}(D)$~\cite{randUint}. From the structure of \erf{dlmpurity} it can be shown that it is not necessary to average over all unitaries; any finite set comprising a unitary 2-design is sufficient~\cite{randUint}. This not only simplifies the calculation below; it will also simplify implementation of this control strategy as only a finite number of different unitaries need be applied.

 We evaluate \erf{Uavg_dlmpurity} using the methods of Ref.~\cite{randUint}. We show this explicitly for the term $T_1\equiv \tr{X U\rho U\dg X U\rho U\dg}$ arising from the first term inside the braces of \erf{dlmpurity}. First we enumerate some preliminary identities.  (i) A permutation operator can be defined, using the Einstein summation convention. For example,  $I= P_{1234}= \op{i}{i}\otimes \op{j}{j}\otimes \op{k}{k}\otimes \op{l}{l}$ and $P_{2341}= \op{i}{l}\otimes \op{j}{i}\otimes \op{k}{j}\otimes \op{l}{k}$, where the indices run from $0$ to $D-1$. (ii) Using this, $\tr{ABCD}=\tr{(A\otimes B\otimes C \otimes D)P_{2341}}$. (iii) Defining $Q'=\int dU\, U\otimes U\dg\otimes U \otimes U\dg$, we find (applying the necessary permutations to Eq. (5.17) of Ref. \cite{randUint}) that $Q'=\left [ D(P_{2143}+P_{4321}) -(P_{4123}+P_{2341})\right ]/D(D^2-1)$. 

Returning to \erf{Uavg_dlmpurity}, the required integral of $T_1$ is 
\begin{eqnarray}\label{terminator1}
\int dU\, T_1
&=& \tr{(X\otimes \rho \otimes X\otimes \rho).Q'. P_{2341}} \\
&=& \mathrm{Tr}\left [ (X\otimes \rho \otimes X\otimes \rho).\left \{D(P_{1432}+P_{3214})\right .\right. \nn\\
&&\phantom{\mathrm{Tr} [}\left.\left.-(P_{1234}+P_{3412})\right\} \right ]/{D(D^2-1)}.
\end{eqnarray}
As an example, we show how to evaluate the second term from this expression: $\mathrm{Tr}\left [ (X\otimes \rho \right . \otimes \left .X\otimes \rho).(DP_{3214})\right ]=D\expt{k|X|i}\expt{j|\rho|j}\expt{i|X|k}\expt{l|\rho|l}=D\tr{X^2}$. Calculating the remaining terms in \erf{terminator1} in like manner gives
\begin{eqnarray}
\int dU\, T_1&=&\tr{X^2}\left(D-\tr{\rho^2}\right)/D(D^2-1).
\end{eqnarray}
A similar calculation must be performed for all terms in \erf{Uavg_dlmpurity}
. The final result is 
\beq\label{exact_dl} \nn
\mathbb{E}[\enavg{dL}]= -\frac{8\tr{X^2}}{(D^2-1)}\gamma dt \left \{ 1-2\tr{\rho^3}+\tr{\rho^2}^2\right\}. 
\eeq
This differential equation cannot be solved exactly because the right-hand-side contains higher-order moments of $\rho$. However we are interested in the the long-time limit, when $\rho$ is very pure. 
Defining the largest eigenvalue of $\rho$ to be $1-\Delta$, we have $L \sim 2\Delta$  and also $B\sim 2\Delta$, where $B$ is the factor in braces on the right-hand-side. Thus 
for small target impurities we have 
\begin{eqnarray}\label{lt_dl}
 \mathbb{E}[\enavg{dL}]&\sim& -\frac{2}{3}D\phantom{.}\gamma dt \left \{ 2\Delta  \right\}\sim-\frac{2}{3}D\phantom{.}\gamma dt L,
\end{eqnarray}
where we have chosen $X=J_z$ as in Ref.~\cite{ComJac0601}, giving $\tr{X^2} = D(D^2-1)/12$. In fact, for all $\rho$, $L\leq B$, so that  $\mathbb{E}[\enavg{L}]$ is bounded above by $\exp(-\frac{2}{3}Dt)L(0)$. 

Now we compare this strategy to continuous measurement without control, in order to quantify the improvement it provides. Our figure of merit is the speed-up: the time it takes the no-control $\enavg{L}$ to reach a fixed impurity divided by the time it takes the above $\mathbb{E}[\enavg{L}]$ to reach that same impurity~\cite{Jac0303}. For the no-control case, $\enavg{L(t)}\propto \exp{(-\gamma t)}$ for long times~\cite{ComJac0601}. Thus the asymptotic speed-up is $S = \frac{2}{3}D$. This is almost the same as the $S=\frac23(D+1)$ found for the feedback algorithm of Ref.~\cite{ComJac0601}. We confirm this high purity result with quantum trajectory simulations in \frf{Fig2}. We see that the random unitary strategy quickly approaches the predicted asymptotic speed-up even for unitaries that are applied moderately frequently ($\delta t = 0.01\gamma^{-1}$). Moreover, significant speed-ups are obtained even for $\delta t = \gamma^{-1}$.   

\begin{figure}
\leavevmode \includegraphics[width=\hsize]{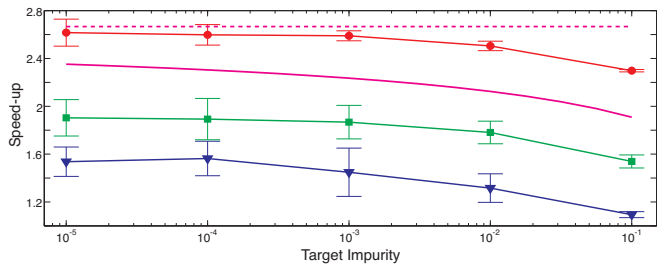}
\caption{(color online). The purification speed-up provided by the random unitary strategy for $D=4$. The  solid curve is a lower bound on the speed-up in the limit where the time $\delta t$ between unitary controls
goes to zero, as explained below \erf{lt_dl}, and the dashed line is its asymptotic (high-purity) limit.   The circles, squares and triangles are numerical calculations of the speed-up with finite $\delta t = 0.01$, $0.25$ and $1$ (in units of the reciprocal measurement rate) respectively. 
}\label{Fig2}
\end{figure}

To conclude this section, we explain how this protocol can be modified for state preparation. After purifying a state to a desired level of purity by the method above we apply one more unitary. This unitary is necessarily a conditional unitary, as it must take the post-purification state to the final target state. In effect we have delayed the feedback to the final unitary. 

\section{2nd application: Rapid Measurement} Many quantum information tasks involve measurement, including readout, measurement-based computation, error correction, and tomography. This motivates 
our second control goal: to minimize the average time $\an{T}$ required to find out which eigenstate of the observable $X$ the system occupied at time $t_0$, with a given confidence \cite{ComWisJac08}. For a control strategy that maintains $\rho$ in the measurement basis, it was shown in Ref.~\cite{ComWisJac08} that a good proxy measure of the asymptotic reduction in $\an{T}$ is the reciprocal of the speed-up, as defined in the preceding section, but with the impurity $L$ replaced by the log infidelity $\ln{(\Delta)}$ (where $\Delta$ is as defined above). In Ref.~\cite{ComWisJac08} a speed-up of  $\Theta(D^2)$ \footnote{The notation $X=\Theta(Y)$ means $X$ is both upper and lower bounded by quantities scaling as $Y$.} was 
shown for $X=J_z$ using control-unitaries drawn from the permutation group ${\mathfrak P}(D)$ (in the measurement basis) according to a locally optimal feedback algorithm. Here we again eschew feedback, and consider an open-loop unitary $U(t)$ given by $U_q\ldots U_2U_1$ for $t_{q}\leq t<t_{q+1}$, where for all $q$, $U_q \in {\mathfrak P}(D)$. 

To obtain an analytical approximation for the speed-up we follow the method used above of treating $\delta t$ as infinitesimal. Explicitly averaging over all permutations, the increment $\mathbb{E}[d\rho]$ is given by  
\beq \label{fbperms}
\sum_{P\in {\mathfrak P}(D) } \cu{\frac{2\gamma}{D!} dt\D{PJ_z P}+\sqrt{\frac{2\gamma}{D!}}\,dw^{(P)}\Hc{PJ_zP} }\rho, \nn
\eeq
where we have defined $D!$ independent Wiener increments.
It is straightforward to show that this equals $\sum^{D-1}_{i=0}\cu{2\gamma dt\aleph\D{\Pi_i} +\sqrt{2\gamma\aleph}dw_i\Hc{\Pi_i}}\rho$, where $\Pi_i = \op{i}{i}$ and $\aleph = D(D+1)/12$. The equation of motion for the populations in the measurement basis is thus 
\begin{eqnarray}
 \mathbb{E}[dp_i]&=&2\sqrt{2\gamma \aleph}\{dw_i(p_i -p_i^2) -p_i\sum_{j\neq i}p_jdw_j\}.
\end{eqnarray}
Taking $p_0$ to be the largest population,  we use the \ito calculus to calculate the average change in the log-infidelity: 
\begin{eqnarray}\label{dlndelta}
\enavg{ \mathbb{E}[d\ln{(\Delta)}]}&=&-(dp _0)^2/2(1-p_0)^2.
\end{eqnarray}
This is easily evaluated using $dw_i\times dw_j= \delta_{ij} dt$. Now we invoke the arguments found in Ref.~\cite{ComWisJac08} to derive lower and upper bounds on this expression. The upper bound is calculated by substituting in the maximally mixed state for a fixed infidelity, $\rho_\mathrm{F}=\mathrm{diag}(1-\Delta,\smallfrac{\Delta}{D-1},\ldots,\smallfrac{\Delta}{D-1})$: 
\begin{eqnarray}
\enavg{ \mathbb{E}[d\ln{(\Delta)}]}_\mathrm{F}
&=& -4dt\gamma \aleph p_0^2D/(D-1).
\end{eqnarray}
The lower bound comes from the minimally mixed distribution for a fixed infidelty,  $\rho_2=\mathrm{diag}(1-\Delta,\Delta,0,\ldots,0)$: 
\begin{eqnarray}
\enavg{ \mathbb{E}[d\ln{(\Delta)}]}_\mathrm{2}&=& -8\gamma \aleph dt p_0^2.
\end{eqnarray}
Consider the asymptotic limit $\Delta \to 0$. For the no-control case, $\enavg{d\ln{(\Delta)}}=-4\gamma t$  \cite{ComWisJac08}. Thus  the control-generated speed-up in measurement is bounded by
\begin{eqnarray}\label{sbounds}
\frac{D^2(D+1)}{12(D-1)}\leq S \leq \frac{D(D+1)}{6}.
\end{eqnarray}
That is, we obtain the same $\Theta(D^2)$ speed-up as for  locally optimal feedback \cite{ComWisJac08}. Once again, one can perform a conditional unitary at the end of the filtering to prepare a desired state with high fidelity. Alternatively, because the permutations are calculationally reversible, the high final fidelity implies a high confidence in the result corresponding to the retrodicted eigenstate, which could be used for rapid readout or rapid tomography, for example. 

\begin{figure}
\leavevmode \includegraphics[width=\hsize]{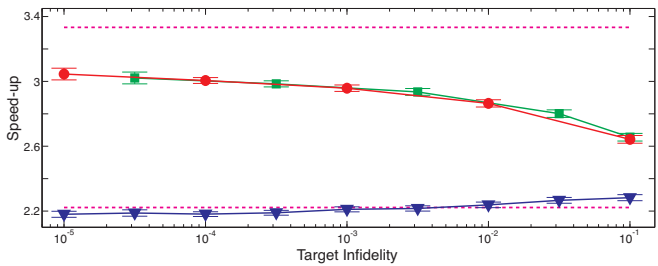}
\caption{(color online). The speed-up for permutation strategies for $D=4$. The dashed lines are the asymptotic bounds in \erf{sbounds}.  The results for random permutation strategy are shown by circles for $\delta t = 6.25\times 10^{-4}\gamma^{-1}$ and triangles for $\delta t= 1.25\times 10^{-3}\gamma^{-1}$. The squares are for the deterministic strategy described in the text with $\delta t = 6.25\times 10^{-4}\gamma^{-1}$.}\label{Fig3} 
\end{figure}

We confirm the bounds in \erf{sbounds} with quantum trajectory simulation in \frf{Fig3}. The numerically calculated speed-up for the random permutation strategy lies within the bounds derived in \erf{sbounds}. We also plot the speed-up for a deterministic control strategy which alternates the permutations  $P_{2143}$ and $P_{3124}$. Remarkably, the speed-up is almost the same as that of the random permutation algorithm. 
Figure \ref{Fig3} also shows that, unlike the random unitary case, the random permutation strategy is quite  sensitive to the frequency of the applied controls. 

We have also considered read-out of a register of qbits, each independently monitored as in Ref. \cite{ComWisJac08}. Numerical results (not shown) indicate that random permutations in the logical basis give a similar improvement to the Hamming-ordered feedback scheme of Ref. \cite{ComWisJac08}.

In summary, despite the feedback delay problem recent theoretical efforts in quantum control have been focused on optimal feedback control, which is particularly computationally time-consuming. In this paper we have demonstrated the effectiveness of combining open-loop control and quantum filtering to replace feedback control for some problems, thus circumventing the feedback delay problem. In addition to this we note that the general strategy we have introduced is another benchmark against which the performance of feedback protocols should be compared.  How the speed-ups obtained in this Letter would be affected by the constraint of bounded control Hamiltonians is an open question, as is the application of our methodology in other systems and for other measurement models. 

\acknowledgments This work was supported by ARC grant  CE0348250. 

\end{document}